\documentstyle[12pt,epsf,psfig]{article}
\input{epsf}

\def\3{\ss}                                           

\def\jpg#1#2#3  {{ J. Phys. G.} {#1} (#2) #3}
\def\cpc#1#2#3  {{ Comp. Phys. Comm.} {#1} (#2) #3}
\def\pl#1#2#3   {{ Phys. Lett.} {#1} (#2) #3}
\def\EP#1#2#3   {{ Eur.  Phys. J.} {#1} (#2) #3}
\def\prep#1#2#3 {{ Phys. Rep.} {#1} (#2) #3}
\def\prev#1#2#3 {{ Phys. Rev.} {#1} (#2) #3}
\def\prl#1#2#3  {{ Phys. Rev. Lett.} {#1} (#2) #3}
\def\zp#1#2#3   {{ Z. Phys.} {#1} (#2) #3}
\def\nim#1#2#3  {{ Nucl. Instr. and Meth.} {#1} (#2) #3}
\def\np#1#2#3   {{ Nucl. Phys.} {#1} (#2) #3}

\def\kt{k_T}
\def\ETJ{E^{{jet}}_T}
\def\xgo{x_\gamma^{\rm OBS}}

\def\ETAJ{\eta^{jet}}

\def\Q2{{\rm Q}^2}
\def\ptmin{\hat{p}_T^{\rm min}}

\begin{document}

\title {
\begin{flushleft}{\normalsize DESY--00--017} \\
{\normalsize February 2000}\end{flushleft}
\vspace{2.0cm}
\bf\LARGE The $\Q2$ Dependence of Dijet Cross Sections
                   in $\gamma p$ Interactions at HERA}  

\author{ZEUS Collaboration}

\date{}

\maketitle

\begin{abstract}
\noindent

The dependence of the photon structure on the photon virtuality,
$\Q2$, is studied by measuring the reaction $e^+p\rightarrow
e^+\ +\ {\rm jet} +\ {\rm jet} +\ {\rm X}$ at photon-proton
centre-of-mass energies 134 $< W < $ 223 GeV.  Events have been 
selected in the $\Q2$ ranges $\approx$ 0~GeV$^2$, 0.1-0.55 GeV$^2$, 
and 1.5-4.5 GeV$^2$, having two jets with transverse 
energy E$_T^{jet} > $ 5.5 GeV in the final state.
The dijet cross section has been measured as a function of the 
fractional momentum of the photon participating in the hard process, 
$\xgo$. The ratio of the dijet cross section with $\xgo < 0.75$ to 
that with $\xgo > 0.75$ decreases as $\Q2$ increases.
The data are compared with the predictions of NLO pQCD and 
leading-order Monte Carlo programs using various parton distribution 
functions of the photon. The measurements can be interpreted in terms 
of a resolved photon component that falls with $\Q2$ but remains present 
at values of $\Q2$ up to 4.5 GeV$^2$. However, none of the models 
considered gives a good description of the data.

\end{abstract}

\pagestyle{plain}
                   
\thispagestyle{empty}
\newpage 
\pagenumbering{Roman}
%
%
%
%
\begin{center}                                                    
{                      \Large  The ZEUS Collaboration              }                               
\end{center}                                                                                       
  J.~Breitweg,                                                                                     
  S.~Chekanov,                                                                                     
  M.~Derrick,                                                                                      
  D.~Krakauer,                                                                                     
  S.~Magill,                                                                                       
  B.~Musgrave,                                                                                     
  A.~Pellegrino,                                                                                   
  J.~Repond,                                                                                       
  R.~Stanek,                                                                                       
  R.~Yoshida\\                                                                                     
 {\it Argonne National Laboratory, Argonne, IL, USA}~$^{p}$
% ...
\newpage                                                                                           
% ...
$^{\    2}$ also at IROE Florence, Italy \\                                                        
% ...
                                                           %
                                                           %
\newpage   
                                                           %
                                                           %
\begin{tabular}[h]{rp{14cm}}                                                                       
$^{a}$ &  supported by the Natural Sciences and Engineering Research                               
          Council of Canada (NSERC)  \\                                                            
$^{b}$ &  supported by the FCAR of Qu\'ebec, Canada  \\                                            
$^{c}$ &  supported by the German Federal Ministry for Education and                               
          Science, Research and Technology (BMBF), under contract                                  
          numbers 057BN19P, 057FR19P, 057HH19P, 057HH29P, 057SI75I \\                              
$^{d}$ &  supported by the MINERVA Gesellschaft f\"ur Forschung GmbH, the                          
German Israeli Foundation, and by the Israel Ministry of Science \\                                
$^{e}$ &  supported by the German-Israeli Foundation, the Israel Science                           
          Foundation, the U.S.-Israel Binational Science Foundation, and by                        
          the Israel Ministry of Science \\                                                        
$^{f}$ &  supported by the Italian National Institute for Nuclear Physics                          
          (INFN) \\                                                                                
$^{g}$ &  supported by the Japanese Ministry of Education, Science and                             
          Culture (the Monbusho) and its grants for Scientific Research \\                         
$^{h}$ &  supported by the Korean Ministry of Education and Korea Science                          
          and Engineering Foundation  \\                                                           
$^{i}$ &  supported by the Netherlands Foundation for Research on                                  
          Matter (FOM) \\                                                                          
$^{j}$ &  supported by the Polish State Committee for Scientific Research,                         
          grant No. 112/E-356/SPUB/DESY/P03/DZ 3/99, 620/E-77/SPUB/DESY/P-03/                      
          DZ 1/99, 2P03B03216, 2P03B04616, 2P03B03517, and by the German                           
          Federal Ministry of Education and Science, Research and Technology (BMBF)\\              
$^{k}$ &  supported by the Polish State Committee for Scientific                                   
          Research (grant No. 2P03B08614 and 2P03B06116) \\                                        
$^{l}$ &  partially supported by the German Federal Ministry for                                   
          Education and Science, Research and Technology (BMBF)  \\                                
$^{m}$ &  supported by the Fund for Fundamental Research of Russian Ministry                       
          for Science and Edu\-cation and by the German Federal Ministry for                       
          Education and Science, Research and Technology (BMBF) \\                                 
$^{n}$ &  supported by the Spanish Ministry of Education                                           
          and Science through funds provided by CICYT \\                                           
$^{o}$ &  supported by the Particle Physics and                                                    
          Astronomy Research Council \\                                                            
$^{p}$ &  supported by the US Department of Energy \\                                              
$^{q}$ &  supported by the US National Science Foundation                                          
\end{tabular}                                                                                      
                                                           %
                                                           %
\newpage
\pagenumbering{arabic}

\section{Introduction}

The photon at high virtuality, $\Q2$, is commonly considered to be a
point-like probe of the structure of a particular hadronic
target \cite{aharon}. However, the real photon ($\Q2 \approx 0$ GeV$^2$)
has itself a partonic structure, which has been studied in two-photon 
reactions from $e^+e^-$ scattering \cite{general}, and in jet production at
HERA \cite{prevpap,saunders}. In this paper, the transition between the
real photon and the virtual photon is investigated for
0 $\leq \Q2 \leq$ 4.5 GeV$^2$ using dijet events in $ep$
scattering at HERA.

The photon, in general, may have both a partonic structure and a 
point-like coupling to charged quarks and leptons. As a result,
two types of process can contribute to jet production in 
$\gamma p$ interactions in leading order (LO) perturbative QCD (pQCD): 
the direct process, in which the 
photon couples directly to quarks at high transverse momenta, one of
which scatters from a parton in the proton, and the resolved process, 
where a parton from the photon scatters from a parton in the proton.
Conventionally, two types of resolved photon process are defined. 
In the first, the photon acts via an intermediate meson-like 
hadronic state whose description is essentially non-perturbative, so 
that a phenomenological parton density function must be introduced.  
In the second, the photon interacts initially by splitting into a 
$q\bar{q}$ pair at moderate transverse energy, a point-like perturbative 
process which is termed `anomalous' and can in principle, for 
$\Q2 >$ 0 GeV$^2$, be summed to all orders. The boundary between the two 
types of resolved process is factorisation-scale dependent.

At a given photon virtuality, $\Q2$, and hard QCD scale, $\mu^2$,  
both types of resolved process can in principle occur.  
It is usually accepted that, at low $\Q2$, the hadronic type is 
important, while at higher $\Q2$, resolved processes are dominated 
by the anomalous type. The general expectation is that the contribution 
to the dijet cross section from both types of resolved photon processes 
should decrease relative to the contribution from direct photon 
processes as the virtuality of the photon increases towards $\mu^2$,
{\it i.e.} the partonic content of the photon becomes
suppressed~\cite{SaS,DG,theory,predictions}.
The first measurements came from the PLUTO collaboration~\cite{PLUTO}.
The H1 collaboration has also studied the transition between
photoproduction and deep inelastic scattering by measuring, in the 
$\gamma^*p$ CM frame, inclusive jet cross sections for real and
virtual photons~\cite{h1_virt_incl} and dijet cross 
sections~\cite{h1_virt_dijet} for $\Q2 > 1.6$ GeV$^2$.

The resolved and direct components can be separated on the basis of the
variable $\xgo$, which is the fractional momentum of the photon 
partaking in the production of the dijet system. 
This variable is defined as:

\[
\xgo = \frac{ \sum_{jets}E_T^{jet}e^{-\eta^{jet}}}{2yE_e}
\]
where $\ETJ$ and $\ETAJ$
are the transverse energy  and pseudorapidity of the jet 
defined in the laboratory frame\footnote{The ZEUS right-handed 
coordinate system is defined with the origin at the nominal 
interaction point by the $Z$ axis pointing in the proton beam 
direction and the $X$ axis pointing horizontally towards the 
centre of HERA.}. The variable $y$ is defined as
$y= 1-\frac{E_e^{'}}{2E_e}(1-\cos{\theta}^{'}_{e})$, where
$E_e$ is the positron beam energy and $E_e^{'}$, ${\theta^{'}_e}$ 
are the energy and polar angle, respectively, of the scattered positron.
Since $\xgo$ is well defined at all orders in pQCD, measurements based 
on it can be compared with theoretical predictions at any given order.

At $\xgo > $ 0.75, the direct component dominates, while the
$\xgo < $ 0.75 region is sensitive mainly to the resolved component.
However, events with low values of $\xgo$ can also be produced when 
initial- and final-state parton showers give rise to hadronic 
activity outside the dijet system.

Experimental $\xgo$ distributions obtained from 1995 ZEUS data are
presented in this paper. The ratio of the measured cross sections
for $\xgo < $ 0.75 and $>$ 0.75 is then given as a function of $\Q2$.
The values of the ratio are compared with theoretical
calculations at both LO and next-to-leading-order (NLO) pQCD
computed using the JetViP program \cite{jetvip}.

\section{Experimental Setup and Data Selection}

During 1995, HERA operated with protons of energy $E_p = 820$~GeV
and positrons of energy $E_e = 27.5$~GeV. The ZEUS detector is
described in detail elsewhere \cite{sigtot,status}. The main components
used in the present analysis are the uranium-scintillator sampling
calorimeter (CAL) \cite{main}, the beam pipe calorimeter (BPC)
\cite{BPCF2}, and the central tracking chamber \cite{CTD}
positioned in a 1.43~T solenoidal magnetic field.
The CAL energy resolution for positrons, under test beam conditions,
was measured to be 0.18/$\sqrt{E_e^{'}({\rm GeV})}$. 
The point of impact of the positron in CAL was measured with a resolution 
of 3 mm, resulting in a Q$^2$ resolution of 8$\%$. The systematic 
uncertainty on the absolute value of $E_e^{'}$ is 2$\%$.
The BPC was installed  294 cm from the interaction point in the 
positron direction in order to tag scattered positrons at small 
angles (15-34 mrad).
It measured both the energy, $E^{'}_e$, of the scattered
positron and the position of its impact point.
The energy resolution of the BPC is 0.17/$\sqrt{E^{'}_e({\rm GeV})}$ and the
position resolution is 0.5 mm, resulting in a $\Q2$ resolution of 6$\%$. 
The systematic uncertainty on the absolute value of $E_e^{'}$ is 0.5$\%$.

The events were selected online via a three-level trigger
system \cite{status,zeuscc94,zeushix} using the same selection 
algorithms
as in a previous dijet publication \cite{saunders}, except that in the
third-level trigger (TLT) the events
were required to have at least two jets with $E_T^{TLT} > 4.0$~GeV and
$\eta^{TLT} < 2.5$. The sample was separated offline into subsamples
corresponding to three different Q$^2$ ranges:

\begin{itemize}

\item Events with quasi-real photons
      ($\Q2 \approx$ 0 GeV$^2$, named PHP in the following)
      were selected by requiring that no identified positron
      was found in the CAL with energy
      $E_e^{'} > $ 5 GeV~\cite{sinistra} and $y <$ 0.7.
      The resulting sample had $\Q2 <1.0$ GeV$^2$ with an estimated
      median of $10^{-3}$ GeV$^2$;

\item   Events at intermediate $\Q2$ (IQS) were selected by
        requiring that the scattered positron was measured by the BPC.
        In this data set, the BPC tagged events with
        photon virtualities in the range 0.1 $ < \Q2 <$ 0.55 GeV$^2$.
        For this sample the energy of the scattered positron
        was required to be $E_e^{'} > $ 12.5 GeV;

\item   Deep inelastic scattering (DIS) events at low $\Q2$ (LDIS)
        were selected by requiring that the outgoing positron was
        measured in the CAL. The energy of the scattered positron
        was required to satisfy $E_e^{'} > $ 11.0 GeV~\cite{sinistra}
        and the Q$^2$ range was restricted to
        1.5 $<$ $\Q2$ $<$ 4.5 GeV$^2$.

\end{itemize}

For all three samples, additional cuts of $0.15 < y_{JB} < 0.45$ were
applied, where $y_{JB}$ is an estimator\footnote{ $y_{JB} = \sum_i
(E_i - E_{Zi}) /2E_e$, where $E_{Zi} = E_i \cos\theta_i$ and $E_i$ is
the energy deposited in the CAL cell $i$ which has a polar
angle $\theta_i$ with respect to the measured $Z$-vertex of the event.
The sum runs over all CAL cells excluding those associated with
a detected scattered positron.} of $y$~\cite{YJB}.
Due to the energy lost in the inactive material in front of the CAL and 
to particles lost in the rear beampipe, $y_{JB}$ systematically 
underestimates the true $y$ by approximately 20\%, an effect which is 
adequately reproduced in the Monte Carlo simulation of the detector.
The combination of $y_{JB}$ and $E_e^{'}$ cuts ensured that all three
samples corresponded to the same true $y$ range (0.2 $< y < $ 0.55).

The longitudinally invariant $\kt$ algorithm~\cite{kt}, in the mode
described in a previous publication \cite{saunders}, was then applied to
the CAL cells to search for events with two jets in the
final state. In the LDIS sample, the cells associated with the 
positron were excluded from the jet search. The two jets with the 
highest transverse energy were required to have pseudorapidity between 
$-1.125 < \eta^{jet} < 2.2 $ and transverse energy $\ETJ > 5.5 $ GeV.

After all cuts, the jet search resulted in a sample of
58224 dijet events for the PHP sample, 353 dijet events for the
IQS sample and 1172 dijet events for the LDIS sample.
Approximately 10$\%$ of the events in each of the three samples
had three or more jets. The PHP, IQS and LDIS samples correspond to
integrated luminosities of 3.1, 3.3 and 4.9 pb$^{-1}$, respectively.

\section{Data Corrections and Systematics}

The data were corrected for acceptance, smearing and kinematic cuts 
using the HERWIG~5.9~\cite{HERWIG} Monte Carlo (MC) model. 
Leading-order resolved (LO-RES) and direct (LO-DIR) processes were 
generated separately. Resolved photon events were generated not only 
in the PHP, but also in the IQS and LDIS regimes. 
The minimum transverse momentum of the partonic hard scatter 
($\ptmin$) was set to 2.5~GeV.
The GRV LO~\cite{GRV} and the MRSA~\cite{MRSA} sets were used for the
photon and proton parton distribution functions (PDF), respectively.
To simulate possible interactions between the proton and photon remnants 
(`underlying event'), the option of multiparton interactions
(MI)~\cite{MI,BFS} was included for the PHP sample.
It has been shown that the simulation of the underlying event with MI
improves the description of the energy flow around the jet axis for jet
production from quasi-real photon-proton interactions~\cite{saunders}.

The Monte Carlo events were processed through the full ZEUS detector
simulation using the same cuts as applied to the data.
The normalisations of the LO-RES and LO-DIR processes were extracted
from the data using a two-parameter fit to the uncorrected $\xgo$
distributions. This procedure was applied separately for each $\Q2$ 
range.

\begin{figure}[p]
\begin{picture}(150,150)(1,150)
\thicklines
\psfig{figure=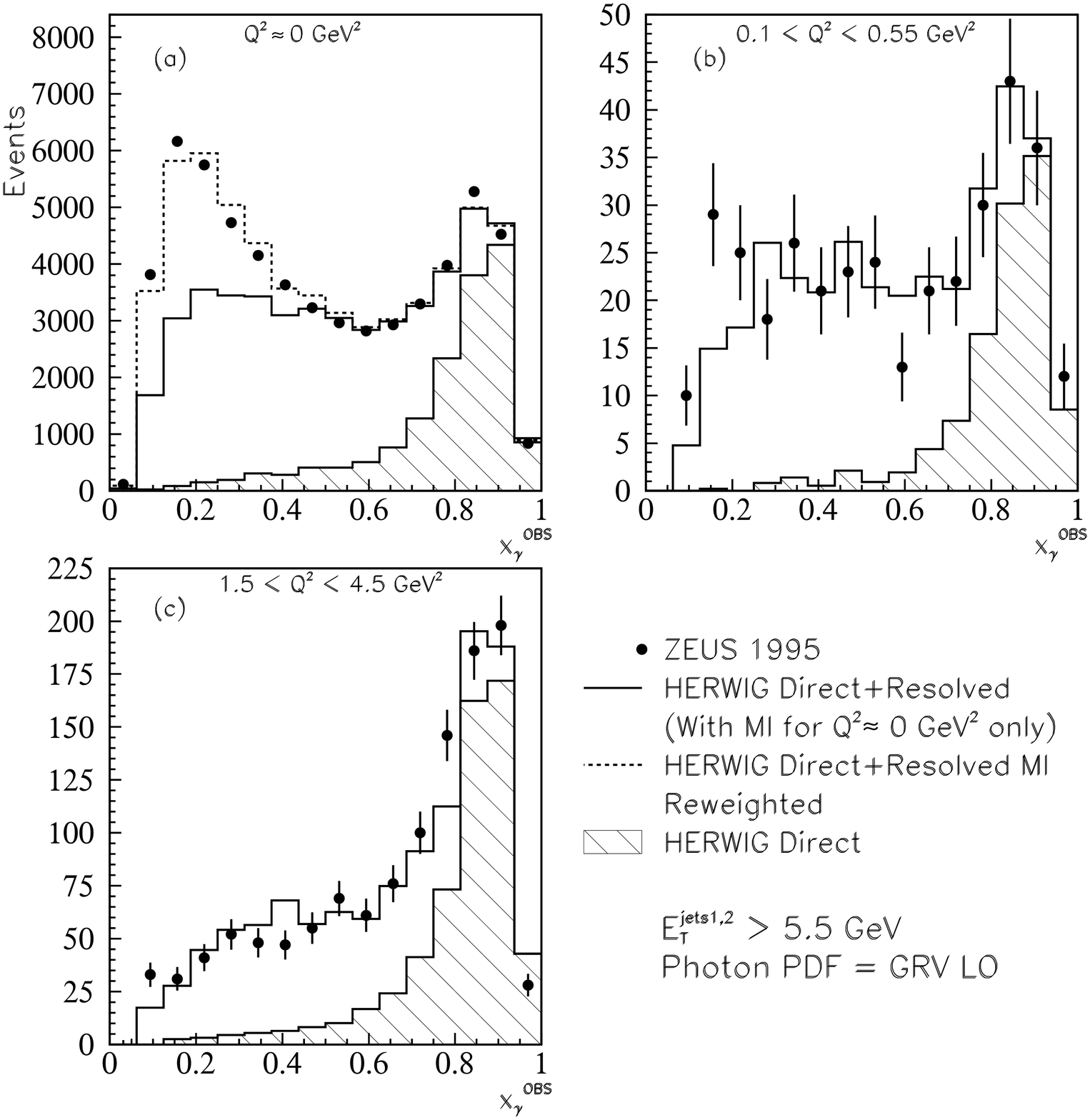,height=15.0cm,width=15.0cm}
\end{picture}
\vspace{5.0cm}
\caption{Uncorrected $\xgo$ distributions for
dijet events selected with the $\kt$ algorithm, for
$\ETJ >$ 5.5 GeV and $-1.125 < \ETAJ < 2.2$, in the ranges:
(a) Q$^2 \approx$ 0 GeV$^2$, (b) 0.1$<$ Q$^2$ $<$ 0.55 GeV$^2$,
and (c) 1.5$<$ Q$^2$ $<$ 4.5 GeV$^2$.
The points are the measurements and the solid histograms are the
predictions of the HERWIG Monte Carlo. The simulation of multiple
parton interactions was used only for the Q$^2 \approx$ 0 GeV$^2$
sample.
In (a) the reweighted predictions of HERWIG (dashed histogram)
used for the data correction are also shown. The shaded histograms
represent the LO-DIR contributions.}
\label{fig:DATA_MC}
\end{figure}

Figure~\ref{fig:DATA_MC} shows uncorrected distributions of $\xgo$ for
PHP, IQS, and LDIS dijet events compared with the HERWIG simulation.
Events both at high $\xgo$, associated mainly with direct photon
processes, and at low $\xgo$, associated mainly with the resolved
photon processes, are present in all $\Q2$ ranges.

In the low $\xgo$ region, the PHP data disagree with the simulation.
Disagreement is also observed in the $\ETAJ$, $y_{JB}$, and $\ETJ$
distributions (not shown) and can be attributed to the presence of
underlying event effects or uncertainty in the PDFs of the
photon. The underlying event effects are most evident at low
$\xgo$ and low $\ETJ$. For the combination of $\ptmin$ and photon
PDF used here, the simulation of multiparton interactions does not
reproduce the shape of the data in the low $\xgo$ region 
\cite{saunders}.
To take account of this disagreement in the correction of the data for 
migrations and acceptance, the Monte Carlo events have been reweighted
as a function of $\xgo$ at the hadron level so that the distribution 
agrees with the data.
The result of the reweighting is shown in
Fig.~\ref{fig:DATA_MC} (a). After the reweighting,
the MC predictions for the $\eta^{jet}$, $y_{JB}$, and $\ETJ$
distributions also agree well with the data (not shown).

The dijet differential cross sections, $d\sigma/d\xgo$, corrected to
the hadron level, have been measured using the $\kt$ jet algorithm,
in the three $\Q2$ regions with 0.2 $ < y < $ 0.55. The measurements
have been made for two sets of jet transverse energy  and pseudorapidity 
cuts:

\begin{enumerate}

\item \underline{Low $\ETJ$}: $\ETJ > 5.5$ GeV, $-1.125 < \ETAJ <
2.2$ for both jets;

\item \underline{High $\ETJ$}: $E_T^{jet_1} > 7.5$ GeV,
$E_T^{jet_2} > 6.5$ GeV, $-1.125 < \ETAJ < 1.875$. 

\end{enumerate}

The data with the low set of $\ETJ$ cuts are sensitive to the
resolved photon component, but also to the effects of the underlying event.
The data with the high $\ETJ$ cuts are not significantly influenced by the
underlying event effects; this was established by means of 
a comparison (not shown) of the data with HERWIG without MI.
The high $\ETJ$ cuts were chosen to be asymmetrical to facilitate 
a comparison with the NLO pQCD calculation.

The cross sections at hadron level were obtained by applying a
bin-by-bin correction to the measured dijet distributions binned
in four $\xgo$ bins
(0.0625-0.25, 0.25-0.50, 0.50-0.75, 0.75-1.00) and
four $\Q2$ bins (0.-1.0, 0.1-0.55, 1.5-3.0, 3.0-4.5 GeV$^2$).
The correction factors take into account the efficiency of the trigger,
the selection criteria and the purity and efficiency of the jet
reconstruction. The efficiency and purity are determined as a function
of $\xgo$ and $\Q2$ from the MC simulation~\cite{BFKL}.
In the PHP region, the correction factors lie between 1.25-1.43 and the 
purities between 0.50-0.64.
In the IQS region, the correction factors are dominated by the
BPC geometric acceptance and lie between 17.6-23.0 and
the purities between 0.40-0.70. 
For the LDIS region the correction factors lie between 3.1-3.8, and
the purities between 0.45 and 0.80.

A detailed study of the systematic uncertainties of the measurements
has been performed \cite{Sean,Neil}. The uncertainties have been
separated into those that are uncorrelated and therefore were added
in quadrature to the statistical error and those that are correlated
and presented separately. The uncorrelated systematic uncertainties
originate from the residual uncertainties in the event simulation.
The uncertainty associated with the $\ETJ$ cut is the  dominant 
uncorrelated
uncertainty for the PHP and IQS samples. When this cut is varied by
the $\ETJ$ resolution of $14\%$, a systematic uncertainty
between $-9\%$ and $+12\%$ results, except for $\xgo < 0.25$ in the
IQS region where the uncertainty ranges between $-28\%$ and $+10\%$. 
In the LDIS region, the dominant systematic uncertainty comes from the
uncertainty in $\xgo$ (the $\xgo$ resolution is 0.05) and results in
a systematic error between $-2\%$ and $+6\%$, except for $\xgo < $ 0.25
where the systematic uncertainty ranges between $-25\%$ and $+36\%$. 

Two sources of correlated systematic uncertainties have been studied, one
originating from the uncertainty of the CAL energy scale and the other
from the use of different models for the description of the jet
fragmentation process in the MC.
The absolute energy scale of the jets in simulated
events has been varied by $\pm 5$\%  \cite{jbjtcg}. The effect of this
variation on the dijet cross sections is $\approx\pm 20$\%.
The uncertainty associated with the jet fragmentation was studied by
correcting the data to the hadron level using PYTHIA~\cite{PYTHIA}
and comparing to the results obtained using HERWIG. The effect
was estimated to be on average $\sim$ 20$\%$. In addition, there is an
overall normalisation uncertainty of 1.5\% from the luminosity
determination, which is not included.

\section{Results and Discussion}

The $\xgo$ distributions shown in Fig.~\ref{fig:DATA_MC},
in all three $\Q2$ ranges, cannot be described by HERWIG without 
including a significant LO resolved photon component, which is 
dominant for $\xgo < 0.75$. Hence the dijet cross sections in 
this region are sensitive to the photon structure.

\begin{figure}[p]
\begin{picture}(150,150)(1,150)
\thicklines
\psfig{figure=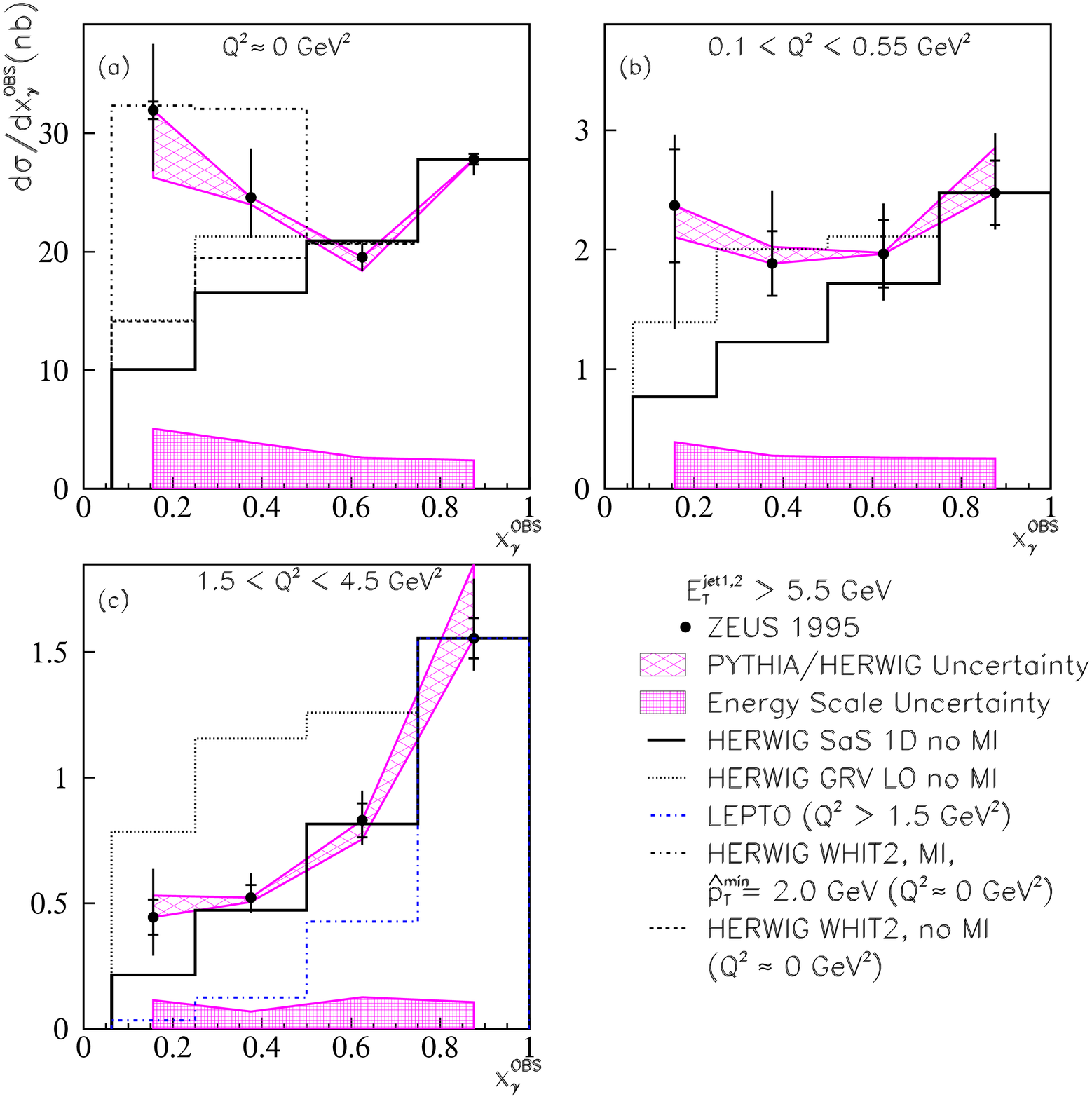,height=14.0cm,width=14.0cm}
\end{picture}
\vspace{5.0cm}
\caption{Dijet cross sections, $d\sigma/d\xgo$, for jets of hadrons
selected with the $\kt$ algorithm in the Q$^2$ ranges:
(a) Q$^2 \approx $ 0 GeV$^2$,
(b) 0.1$<$ Q$^2$ $<$ 0.55 GeV$^2$,
(c) 1.5$<$ Q$^2$ $<$ 4.5 GeV$^2$ for the low $\ETJ$
set of cuts. The points represent the measured cross sections.
The inner error bars represent the statistical errors, and the outer
are the statistical and uncorrelated systematic errors added in quadrature.
The shaded band represents the systematic uncertainty due to the
modelling of the jet fragmentation, estimated using PYTHIA.
The shaded horizontal band represents the uncertainty due to the CAL 
energy scale.
The full histogram represents the HERWIG predictions without MI
using the SaS 1D photon PDFs, and the dotted histogram represents
those with the GRV LO real photon PDFs.
The predictions of HERWIG without MI using the WHIT2 set 
(dashed histogram) and with MI for $\ptmin = $ 2.0 GeV
(dot--dashed histogram) are shown in (a). The predictions of LEPTO are
shown in (c) as the dot-dashed histogram. The LEPTO and HERWIG 
predictions
have been normalised to the $\xgo >$ 0.75 dijet cross section.}
\label{fig:CROSS_SECTIONS2}
\end{figure}

\begin{figure}[p]
\begin{picture}(150,150)(1,150)
\thicklines
\psfig{figure=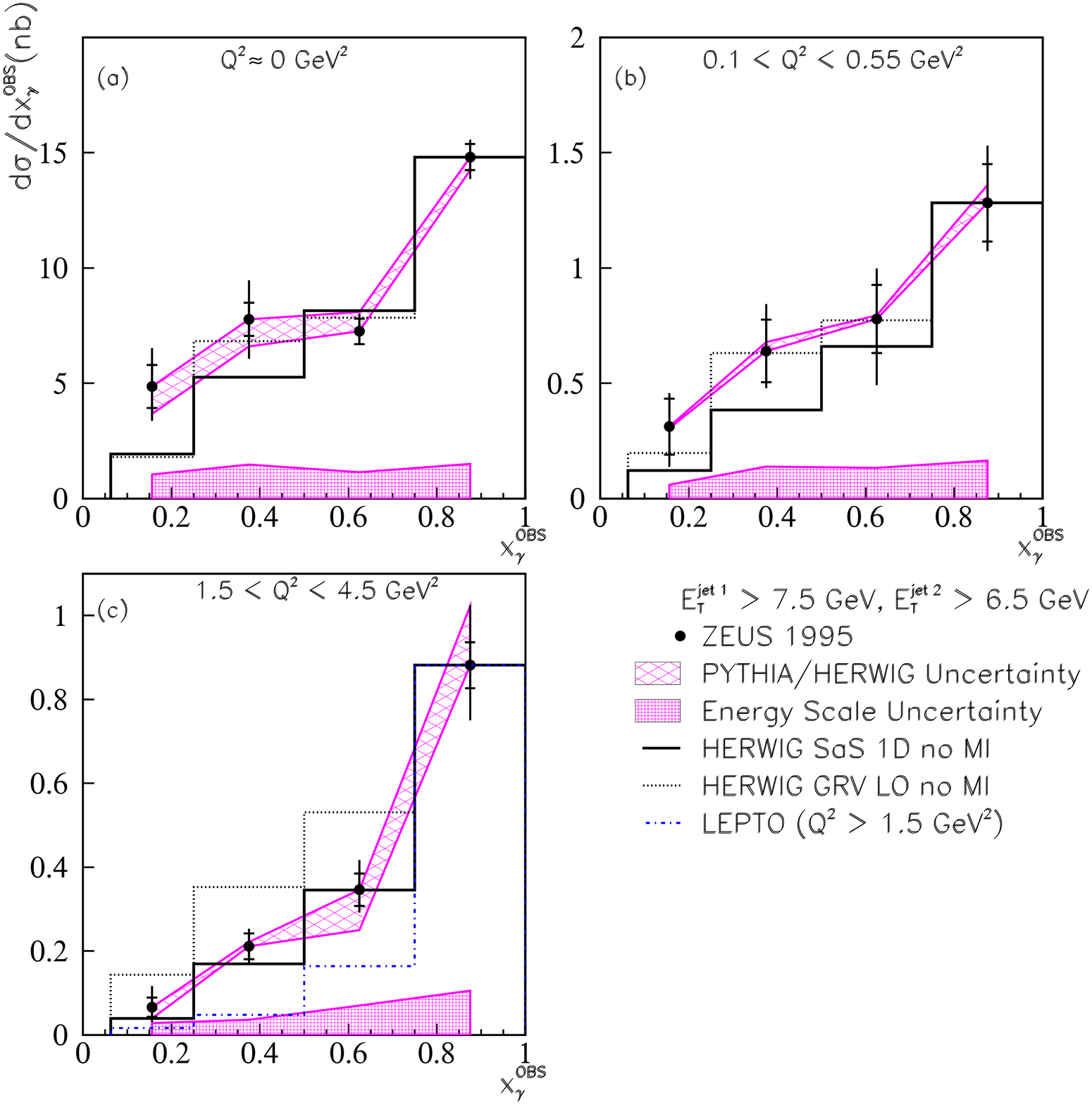,height=14.0cm,width=14.0cm}
\end{picture}
\vspace{5.0cm}
\caption{Dijet cross sections, $d\sigma/d\xgo$, for jets of hadrons
selected with the $\kt$ algorithm in the Q$^2$ ranges :
(a) Q$^2 \approx $ 0 GeV$^2$,
(b) 0.1$<$ Q$^2$ $<$ 0.55 GeV$^2$,
(c) 1.5$<$ Q$^2$ $<$ 4.5 GeV$^2$ for the high $\ETJ$ set of
cuts. The points represent the measured cross sections.
The inner error bars represent the statistical errors and the outer
are the statistical and the uncorrelated systematic errors added in 
quadrature.
The shaded band represents the systematic uncertainty due to the 
modelling of the jet fragmentation, estimated using PYTHIA.
The shaded horizontal band represents the uncertainty due to the
CAL energy scale.
The full histogram represents the HERWIG prediction 
without MI
using the SaS 1D photon PDFs, and the dotted histogram the GRV LO photon 
PDFs. The predictions of LEPTO are shown
in (c) as the dot-dashed histogram. The LEPTO and HERWIG predictions have
been normalised to the $\xgo >$ 0.75 dijet cross section.}
\label{fig:CROSS_SECTIONS1}
\end{figure}

The measured dijet cross sections for the low and the high $\ETJ$
cuts described in Section~3 are shown in Figs.~\ref{fig:CROSS_SECTIONS2}
and \ref{fig:CROSS_SECTIONS1}, respectively.
The shapes of the dijet cross sections change markedly with
increasing $\Q2$, the cross section in the low-$\xgo$ region decreasing 
faster than the cross section in the high-$\xgo$ region. 
This effect is more pronounced for the low $\ETJ$ cuts.

The dijet cross sections are compared to the predictions of the
HERWIG MC at hadron level using different photon PDFs.
Those of  GRV LO~\cite{GRV} and WHIT2~\cite{WHITG2} are valid for
real photons only, have differing gluon distributions and have
no suppression of the resolved photon component as $\Q2$ increases.
In the SaS 1D \cite{SaS} model the resolved photon consists of two
separate contributions, the non-perturbative hadronic 
`Vector Meson Dominance' component and the anomalous pQCD component, 
each with different $\Q2$ dependence. 
Specifically, the `Vector Meson Dominance'
component of the resolved photon is predicted to decrease
approximately as $ (m_{\rho}^2/(m_{\rho}^2+\Q2))^2$. 
The pQCD component is predicted to decrease more
slowly as $\sim \log(\mu^2/\Q2)$, where $\mu^2$ is the hard QCD scale
of the process which, for jet production, is usually taken
to be proportional to ${(\ETJ)}^2$.
The measured cross sections for the LDIS region are also
compared to the LEPTO \cite{LEPTO} Monte Carlo prediction,
which does not include a resolved photon component and uses
a parton-shower model to account for higher-order pQCD effects.
The general framework is similar to the LO-DIR HERWIG and
PYTHIA simulations.  In this picture, the dijet cross section
at low $\xgo$ arises purely from parton-shower contributions
to the LO-DIR process. The HERWIG and LEPTO predictions agree in
the highest $\xgo$ bin, where the direct component dominates.
In order to compare the shape of the measured cross sections
with that of the MC predictions, the latter have been normalized
to the data cross sections for $\xgo > 0.75$.

The low $\ETJ$ cross sections are  compared to the HERWIG 
predictions in Fig.~\ref{fig:CROSS_SECTIONS2}.
The SaS 1D prediction without MI agrees qualitatively with the data
in the LDIS range; however a disagreement is observed at low
$\xgo$ in the IQS and PHP ranges, which becomes more striking as
$\Q2$ decreases.
The GRV prediction without MI and the WHIT2 prediction both without
and with MI using $\ptmin = 2.0$~GeV are compared to the data
in the PHP region, where the discrepancy with SaS 1D is greatest.
The effect of MI is found to be very sensitive to the $\ptmin$ 
value and to the choice of PDF~\cite{BFS,jb99}.  The model using 
WHIT2 with MI gives reasonable agreement with the data.
The shape of the low $\ETJ$-cut cross sections, shown in
Fig.~\ref{fig:CROSS_SECTIONS2} (a), cannot be described by the models
that do not include MI.
The discrepancy seen for the PHP data using SaS 1D without MI is not
present in the LDIS region. This is as expected, in the framework of
the MI model, if the resolved component decreases with $\Q2$. 
The LEPTO predictions underestimate the dijet cross sections at low
$\xgo$ in the LDIS region, indicating that the parton-shower
contributions alone cannot describe the dijet data in this region.

The high-$\ETJ$ data are shown in Fig.~\ref{fig:CROSS_SECTIONS1}.
The predictions of HERWIG without multi-parton
interactions using the SaS 1D photon PDF describe the shape of
the measured cross section well in the LDIS region but tend to
underestimate the PHP and IQS data at low $\xgo$.
The measurements are also compared to HERWIG using GRV without MI.
This model is in good agreement with the data in the PHP and IQS
regions but fails to describe the data in the LDIS region, as
expected since the GRV set describing the real photon structure
is used. As seen in Fig.~\ref{fig:CROSS_SECTIONS1} (c), LEPTO
again underestimates the dijet cross sections at low $\xgo$.

\begin{figure}[p]
\unitlength 1.0cm
\begin{picture}(10,10)(-2.5,0.0)
\thicklines
\psfig{figure=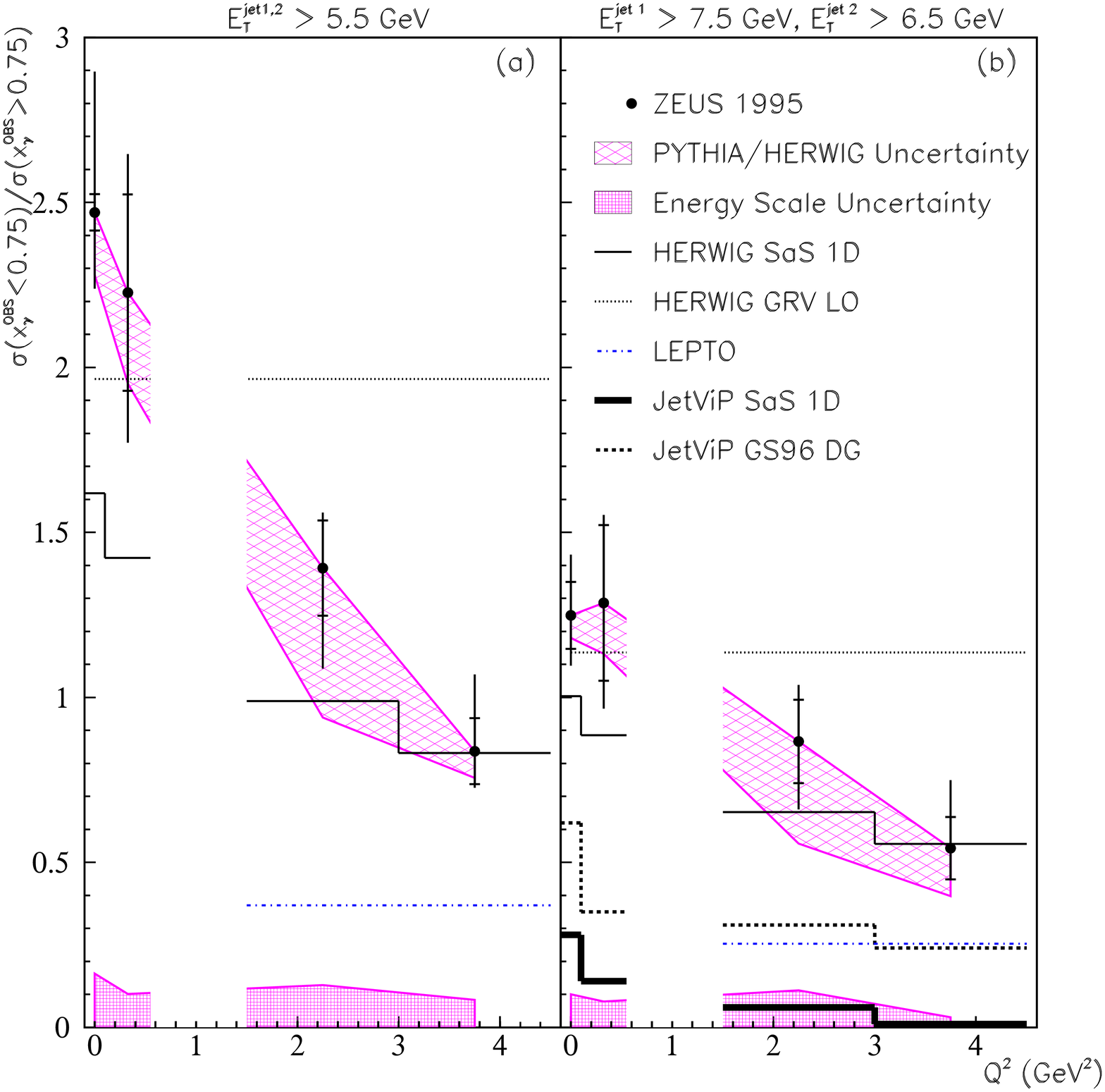,height=5in,width=5in}
\end{picture}
\caption{The ratio of dijet cross sections,
$\sigma(\xgo<0.75)/\sigma(\xgo>0.75)$, 
as a function of photon virtuality, $Q^2$, for dijet events
selected using the $\kt$ algorithm. The points represent
the ZEUS data. The cross section ratios  are shown for both
the low (a) and the high (b) $\ETJ$ cuts.
The inner error bars represent the statistical errors and the
outer are the statistical and systematic errors added in quadrature.
The shaded band represents the systematic error due to the uncertainty
in modelling the jet fragmentation, estimated using PYTHIA.
The shaded horizontal band represents the uncertainty due to the
CAL energy scale. Also shown are the predictions of
HERWIG without MI for two different choices of photon PDFs: GRV for real
photons (dotted histogram) and SaS 1D (full histogram).
The LEPTO predictions are shown for Q$^2 >$ 1.5 GeV$^2$
(dot--dashed histogram). The predictions of NLO pQCD calculated
using the JetViP program with SaS 1D Photon PDFs are shown as the     
full histogram and those using GS96 DG are shown as dashed the histogram.}

\label{fig:RATIOS}
\end{figure}

The cross-section ratio
$\sigma(x_{\gamma}^{\rm OBS}<0.75)/\sigma(x_{\gamma}^{\rm OBS}>0.75)$
as a function of $\Q2$ for both sets of $E_T^{jet}$ cuts is shown
in Fig.~\ref{fig:RATIOS}. The dominant systematic uncertainties
of these measurements (7-16$\%$) are due to the $\ETJ$ and $\xgo$
cuts, except for the LDIS samples where the cut on the impact
point of the scattered positron results in an additional systematic
uncertainty of about 10$\%$.  For the IQS measurements, the latter
systematic uncertainty falls to 5$\%$. 
When the data are corrected using PYTHIA, the measured ratios are
systematically lower for all $\Q2$ points. This systematic error
is therefore not included with the previous ones, but it is shown
separately.
For the PHP data, there is an additional error of
5$\%$ due to uncertainties in the Monte Carlo normalisation factors
for the LO-DIR and LO-RES used in the fit (not shown).

The cross-section ratio falls steeply as a function of $\Q2$.
This can be interpreted as the suppression of the resolved photon
component as the photon virtuality increases. The decrease is more
pronounced for the measurements using the low set of $\ETJ$ cuts,
which are more sensitive to the resolved component and a possible
underlying event. The predictions of HERWIG with two different photon
PDFs are also shown. The prediction using the GRV set is flat,
irrespective of the presence of MI, as expected for a photon PDF
lacking a $\Q2$ dependence. The prediction using the SaS 1D PDF
decreases with $\Q2$ and lies below the data in the low $\Q2$ region.
The measured ratios are also compared with the predictions of LEPTO
in the LDIS region in which this model is applicable.
The LEPTO predictions show the contribution to the ratio arising
from parton shower effects alone and underestimate the measured
ratios in both cases.

In Fig.~\ref{fig:RATIOS}(b), the high-$\ETJ$ data are also compared
to the predictions of a NLO pQCD calculation at the parton level
using the program JetViP \cite{jetvip}. The renormalisation and
factorisation scales were set to Q$^2 + {(\ETJ)}^2$.
The calculation includes contributions from a resolved photon
component, which are computed using two different sets of photon PDFs:
the SaS 1D PDF and the GS96 HO \cite{GS} PDF modified to include a $\Q2$ 
suppression according to Drees and Godbole \cite{DG} (GS96 DG).
The JetViP predictions
are sensitive to the choice of the photon PDFs but lie well
below the data. The magnitude of the hadron-to-parton level 
corrections
has been investigated as a possible source of this discrepancy.
The data corrections to parton level were estimated using the MC
samples and were found to decrease the measured cross section ratios
by approximately 20-30$\%$, which is insufficient to explain the
discrepancy.

\section{Conclusions}

Dijet cross sections, $d\sigma/d\xgo$, have been measured
using the longitudinally-invariant $\kt$ jet algorithm
as a function of $\Q2$, for Q$^2 < 1$ GeV$^2$,
0.1$<$ Q$^2$ $<$ 0.55 GeV$^2$
and 1.5$<$ Q$^2$ $<$ 4.5 GeV$^2$. The $\xgo$ dependence of
the measured dijet cross sections changes with increasing
$\Q2$. The low-$\xgo$ cross section decreases more rapidly
than the high-$\xgo$ cross section as $\Q2$ increases.
This effect is more pronounced for the lower of the two sets of
$\ETJ$ cuts.

The shape of the dijet cross sections, $d\sigma/d\xgo$, is compared to
the predictions of HERWIG MC for a variety of photon PDFs. None of these
models is able to explain the data for both high- and low-$\ETJ$ cuts in
all $\Q2$ ranges.

The ratio $\sigma(\xgo<0.75)/\sigma(\xgo>0.75)$ for dijet cross
sections decreases as $\Q2$ increases but remains above the level
expected from parton-shower effects alone. This may be interpreted
in terms  of a resolved photon component which is suppressed as the
photon virtuality increases but which remains present up to
$\Q2$ = 4.5 GeV$^2$ when the photon is probed at the scale
$\mu^2 \sim $ 30 GeV$^2$ of these measurements.
Within the models available, events at $\xgo <$ 0.75 can originate from
non-perturbative photon structure or perturbatively-calculable
higher-order processes, and are influenced by underlying-event effects
especially at low $\Q2$ and low $\ETJ$. The features and trends seen in 
the data are in accord with general expectations. However, none of the LO 
models, or the NLO calculation examined here, gives a good description 
of the data across the full kinematic region.

\vspace{0.5cm}
\noindent {\Large\bf Acknowledgements}
\vspace{0.3cm}

The design,
construction and installation of the ZEUS detector have been made
possible by the ingenuity and dedicated efforts of many people from
inside DESY and from the home institutes who are not listed as
authors. Their contributions are acknowledged with great
appreciation. The experiment was made possible by the inventiveness and
the diligent efforts of the HERA machine group.
The strong support and encouragement of the DESY directorate have been
invaluable.
We would like to thank B. P\"otter and G. Kramer for valuable 
discussions and for providing the NLO
calculations. We would also like to thank M. Drees, R. Godbole,
B. Harris, M. Klasen and J. Dainton for helpful discussions.

\setcounter{secnumdepth}{0} 

\newpage

\begin{thebibliography}{99}
%

\bibitem{aharon}  H1 Collab., S. Aid et al., \np{B470}{1996}{3};\\
                  ZEUS Collab., M. Derrick et al., 
\zp{C72}{1996}{399};\\
                  ZEUS Collab., J. Breitweg et al., 
\pl{B407}{1997}{432};\\
                  H1 Collab., C. Adloff et al., \np{B497}{1997}{3};\\
                  A. Levy, \pl{B404}{1997}{369}.
%
\bibitem{general} JADE Collab. W. Bartel et al.,   
\zp{C24}{1984}{231};\\
                  TASSO Collab., M. Althoff at al., 
\zp{C31}{1986}{527};\\
                  PLUTO Collab., C. Berger at al., 
\np{B281}{1987}{365};\\
                  TPC/2$\gamma$ Collab. H. Aihara et 
al.,\zp{C34}{1987}{1};\\
                  L3 Collab., M. Acciarri et al.,\pl{B436}{1998}{403};\\
                  ALEPH Collab., R. Barate et al., 
\pl{B458}{1999}{152};\\
                  L3 Collab., M. Acciarri et al., 
\pl{B447}{1999}{147};\\
                  L3 Collab., M. Acciarri et al., \pl{B453}{1999}{333}.
%
\bibitem{prevpap} H1 Collab., T. Ahmed et al., \pl{B297}{1992}{205};\\
                  ZEUS Collab., M. Derrick et al., 
\pl{B297}{1992}{404};\\
                  H1 Collab., I. Abt et al., \pl{B314}{1993}{436}; \\
                  ZEUS Collab., M. Derrick et al., 
\pl{B322}{1994}{287};\\
                  ZEUS Collab., M. Derrick et al., 
\pl{B342}{1995}{417};\\
                  ZEUS Collab., M. Derrick et al., \pl{B348}{1995}{665}; 
\\
                  ZEUS Collab., J. Breitweg et al.,
\EP{C4}{1998}{591};\\
                  H1 Collab., C. Adloff et al., \EP{C1}{1998}{97};\\
                  ZEUS Collab., J. Breitweg et al., \EP{C11}{1999}{35}.
%
\bibitem{saunders} ZEUS Collab., J. Breitweg et al., \EP{C1}{1998}{109}.
%
\bibitem{SaS} G. Schuler and T Sj\"ostrand, \pl{B376}{1996}{193}.
%
\bibitem{DG} M.Drees and R.Godbole, \prev{D50}{1994}{3124}.
%
\bibitem{theory} T. Uematsu and T. Walsh, \pl{B101}{1981}{263};\\
                 T. Uematsu and T. Walsh, \np{B199}{1982}{93};\\
                 F. Borzumati and G. Schuler, \zp{C58}{1993}{139}; \\
                 M. Gl\"uck, E. Reya and I. Schienbein,
\prev{D60}{1999}{54019}.
%
\bibitem{predictions} M. Gl\"uck, E. Reya and M. Stratmann, 
\prev{D54}{1996}{5515};\\
                      D.de Florian, C. Canal and R. Sassot,
\zp{C75}{1997}{265};\\
                      M. Klasen, G. Kramer and B. P\"otter, 
\EP{C1}{1998}{261}.
%
\bibitem{PLUTO} PLUTO Collab., Ch. Berger et al., 
\pl{B142}{1984}{119};\\
                L3 Collab. F. Ern\'e et al., Proc. of Photon 99 Conf.,
                23-27 May 1999,\\ Freiburg, Germany (to appear).
%
\bibitem{h1_virt_incl} H1 Collab., C. Adloff et al., 
\pl{B415}{1997}{418}.
%
\bibitem{h1_virt_dijet} H1 Collab., C. Adloff et al.,
                        DESY-98-205, submitted to Eur. Phys. J. C.
%
%
\bibitem{jetvip} G. Kramer and B. P\"otter, \EP{C5}{1998}{665};\\
                 B. P\"otter, \EP{direct C5}{1999}{1};\\
                 B. P\"otter, \cpc{119}{1999}{4};\\
                 B. P\"otter,  private communication.
%
\bibitem{sigtot} ZEUS Collab., M. Derrick et al., \pl{B293}{1992}{465}.
%
\bibitem{status} The ZEUS Detector, Status Report 1993, DESY 1993.
%
\bibitem{main} M. Derrick et al., \nim{A309}{1991}{77}; \\
               A. Andresen et al., \nim{A309}{1991}{101};\\
               A. Bernstein et al., \nim{A336}{1993}{23}.
%
\bibitem{BPCF2} ZEUS Collab., J.Breitweg et al., \pl{B407}{1997}{432}.
%
\bibitem{CTD}   N.~Harnew et al., \nim{A279}{1989}{290};\\
                B.~Foster et al., \np{B32}{1993}{181} (Proc.~Suppl.);\\
                B.~Foster et al., \nim{A338}{1994}{254}.
%
\bibitem{zeuscc94} ZEUS Collab., M. Derrick et al., \zp{C72}{1996}{47}.
%
\bibitem{zeushix} ZEUS Collab., J. Breitweg et al., \zp{C74}{1997}{207}.
%
\bibitem{sinistra}  H. Abramowicz, A. Caldwell and R. Sinkus, \\
\nim{A365}{1995}{508}.
%
\bibitem{YJB} F. Jacquet and A. Blondel, Proceedings, Study of an $ep$ 
facility \\
               for Europe,  Hamburg, ed. U. Amaldi, DESY 79-48, (1979) 
391.
%
\bibitem{kt}  S. Catani et al.,
              \np{B406}{1993}{187};\\
              S.D. Ellis and D.E. Soper, \prev{D48}{1993}{3160}.
%
\bibitem{HERWIG} G. Marchesini et al., \cpc{67}{1992}{465}.
%
\bibitem{GRV} M. Gl\"uck, E. Reya and A. Vogt, \prev{D46}{1992}{1973}.
%
\bibitem{MRSA} A.D. Martin, W.J. Stirling and R.G. Roberts, 
\prev{D50}{1994}{6734}.
%
\bibitem{MI}T. Sj\"ostrand and M. van Zijl, \prev{D36}{1987}{2019};\\
            G. Schuler and T. Sj\"ostrand, \pl{B300}{1993}{169};\\
            G. Schuler and T. Sj\"ostrand, \np{B407}{1993}{539}; \\
            J.M. Butterworth and J.R. Forshaw, \jpg{19}{1993}{1657}.

\bibitem{BFS} J. M. Butterworth, J. R. Forshaw and M. H. Seymour, 
\zp{C72}{1996}{637}.
%
\bibitem{BFKL} ZEUS Collab., J. Breitweg et al., \EP{C6}{1999}{239}.
%
\bibitem{Sean} Sean Mattingly, `Virtual Photon Structure with ZEUS at 
HERA',\\
               Ph.D Thesis, University of Wisconsin - Madison, (1999) 
(unpublished).
%
\bibitem{Neil} N. Macdonald, `Structure of the Virtual Photon at HERA', 
\\
               Ph.D Thesis, University of Glasgow (1999) (unpublished).
%
\bibitem{jbjtcg} ZEUS Collab., M. Derrick et al., 
\pl{B342}{1995}{417};\\
                 ZEUS Collab., M. Derrick et al., \pl{B348}{1995}{665}.
%
\bibitem{PYTHIA}  T. Sj\"ostrand,  \cpc{82}{1994}{74}.
%
\bibitem{WHITG2}  K. Hagiwara, M. Tanaka and I. Watanabe, 
\prev{D51}{1995}{3197}.
%
\bibitem{LEPTO} LEPTO 6.5, G. Ingelman, A. Edin and J. Rathsman, \\
                \cpc{101}{1997}{108}.
%
\bibitem{jb99} J. Butterworth and R. Taylor , 
              `A global study of photon induced jet production',\\
               hep-ph/9907394, Proc. of Photon 99 Conf., 23-27 May 1999,\\
               Freiburg, Germany (to appear).
%
\bibitem{GS}     L. Gordon and J.K. Storrow, \np{B489}{1997}{405}.
%
\end{thebibliography}
\end{document}